\documentclass[12pt,doublespace]{article}
\begin{document}
\begin{center}
INVERSE CHAOS SYNCHRONIZATION BETWEEN BIDIRECTIONALLY COUPLED VARIABLE MULTIPLE TIME DELAY SYSTEMS\\
E.M.Shahverdiev\\
Institute of Physics, H.Javid Avenue,33, Baku, AZ1143, Azerbaijan\\
e-mail:shahverdiev@physics.ab.az\\
~\\
ABSTRACT\\
\end{center}
We make first report on inverse chaos synchronization between bi-directionally non-linearly and linearly coupled variable multiple time delay Ikeda systems.The results are of certain importance in secure chaos-based communication systems.
~\\
PACS number(s):05.45.Xt, 05.45.Vx, 42.55.Px, 42.65.Sf\\
\begin{center}
1.INTRODUCTION
\end{center}
\indent 
\indent In recent years the presence of chaoic vibrations and its control in nonlinear dynamical systems in physics, power electronics, ecology, economics and so on, has been extensively demonstrated and is now well established,see,e.g. references in [1-2]. Chaos synchronization [3] is another basic feature in nonlinear science and is one of fundamental importance in a variety of complex physical, chemical, and biological systems, see e.g.references in [1-2]. Potential application areas of chaos synchronization include secure communications, optimization of nonlinear system performance,information processing, and pattern recognition phenomena [1-2].\\
\indent Recently, delay differential equations [4] have attracted much attention in the field of nonlinear dynamics. The high complexity of the multiple time-delayed systems can provide a new architecture for enhancing message security in chaos based encryption systems [5]. In such communication systems message decoding would require chaos synchronization between multiple time-delayed transmitter and receiver systems [1-3]. Variable multiple time delay systems are further generalization of the fixed time delay systems [6]. Such variations of time delays could be intentionally or as a result of fluctuations. In a word, modulated time delay systems could be more realistic models of interacting systems. Investigation of synchronization possibilities in such systems are of certain importance.\\
\indent There are different types of sychronization in interacting chaotic systems [1-2]. Complete, generalized, phase, lag and anticipating synchronizations of chaotic vibrations have been described theoretically and observed experimentally. Complete synchronization implies coincidence of states of interacting systems, $y(t)=x(t)$ [3]; a generalized synchronization is defined as the presence of some functional relation between the states of response and drive, i.e. $y(t)=F(x(t))$; phase synchronization means entrainment of phases of chaotic oscillators, $n\Phi_{x}-m\Phi_{y}=const,$ ($n$ and $m$ are integers) whereas their amplitudes remain chaotic and uncorrelated; lag synchronization appears as a coincidence of shifted-in-time states of two systems, $y(t)= x_{\tau}(t)\equiv 
x(t- \tau)$ with positive $\tau$; anticipating synchronization also appears as a coincidence of shifted-in-time states of two coupled systems, but in this case, in contrast to lag synchronization, the driven system anticipates the driver, $y(t)=x(t+\tau)$ or $x=y_{\tau}$,$\tau >0$ [1-2]. An experimental observation of anticipating synchronization in external cavity laser diodes has been reported in [7]. For inverse synchronization [8], a time-delayed chaotic system $x$ is coupled with an another system $y$ in such a way that one system's dynamics synchronize to the inverse state of the other system or vice-versa: $x(t)=-y(t).$\\
\indent Most chaos based communication techniques use synchronization in unidirectional master-slave system.Such a coupling scheme prevents the messages being exchanged between the sender and receiver. A two way transmission of signals requires bidirectional coupling. With this in mind this paper presents the first report of the inverse chaos synchronization between bi-directionally non-linearly and linearly coupled modulated multiple time delayed Ikeda models with two feedbacks.\\
\begin{center}
II. SYSTEM MODEL
\end{center}
\indent First consider inverse synchronization between bi-directionally {\it nonlinearly} coupled variable time-delayed Ikeda systems with two feedbacks,
\begin{equation}
\frac{dx}{dt}=-\alpha x - m_{1} \sin x_{\tau_{1}}
- m_{2} \sin x_{\tau_{2}} + K_{y} \sin y_{\tau_{3}}
\end{equation}
\begin{equation}
\frac{dy}{dt}=-\alpha y - m_{3} \sin y_{\tau_{1}}
- m_{4} \sin y_{\tau_{2}} + K_{x} \sin x_{\tau_{3}}
\end{equation}
\indent This investigation is of considerable practical importance, as the equations of
the class B lasers with feedback (typical representatives of class B are solid-state,
semiconductor, and low pressure $CO_{2}$ lasers) can be reduced to an equation of the
Ikeda type,see e.g.references in [8].The Ikeda model was introduced to describe the dynamics of an optical bistable resonator,playing an important role in electronics and physiological studies and is well-known for
delay-induced chaotic behavior,see e.g.references in [8].
Physically $x$ is the phase lag of the electric field across the resonator; $\alpha$ is the relaxation coefficient for the driving $x$ and driven $y$ dynamical variables; $\tau_{1,2}=\tau_{01,02} + x_{1}(t) \tau_{a1,a2}\sin(\omega_{1,2}t)$ are the variable feedback loop delay times;$\tau_{3}=\tau_{03} + x_{1}(t)\tau_{a3}\sin(\omega_{3}t)$ is the variable time of flight between systems $x$ and $y$;$\tau_{01,02,03}$ are the zero-frequency component,$\tau_{a1,a2,a3}$ are the amplitude,$\frac{\omega_{1,2,3}}{2\pi}$ are the frequency of the modulations; $x_{1}(t)$ is the output of system (1) ($K_{y}=0$) for constant time delays, i.e. $\tau_{1}=\tau_{01}, \tau_{2}=\tau_{02};$ $m_{1,2}$ and $m_{3,4}$ are the feedback strengths for $x$ and $y$ systems, respectively; $K_{x,y}$ are the coupling strengths between the systems. Variable time delays $\tau_{1,2,3}$ are chosen in such a way to include both chaotic and non-chaotic components.\\
Before considering the case of modulated time delays, we present the conditions for inverse synchronization for fixed time delays, i.e.$\omega_{1,2,3}=0.$ \\
One finds that systems (1) and (2) can be synchronized on the inverse synchronization regime 
\begin{equation}
x=-y 
\end{equation}
as the synchronization error signal $\Delta=x-(-y)=x+y$  for small $\Delta$
under the condition
\begin{equation}
m_{1}=m_{3}, m_{2}=m_{4}, K_{x}=K_{y}
\end{equation}
 obeys the following dynamics
\begin{equation}
\frac{d\Delta}{dt}= -\alpha\Delta - m_{3} \Delta_{\tau_{01}} \cos x_{\tau_{01}} - m_{4} \Delta_{\tau_{02}} \cos  x_{\tau_{02}} + K_{y}\Delta_{\tau_{03}} \cos x_{\tau_{03}}
\end{equation}
It is obvious that $\Delta= 0$ is a solution of system (5).It is noted that solution $x=-y$ under the conditions (4) is also follows from the symmetry of Eqs.(1) and (2).\\
One can study the stability of the synchronization regime $x=-y$ by using the 
Krasovskii-Lyapunov 
functional approach. According to [4], the sufficient stability condition for the trivial 
solution $\Delta=0$ of time-delayed 
equation $\frac{d\Delta}{dt}=-r(t)\Delta + s_{1}(t)\Delta_{\tau_{1}}+ s_{2}(t)\Delta_{\tau_{2}} + s_{3}(t)\Delta_{\tau_{3}}$ 
is: $r(t)>\vert s_{1}(t) \vert + \vert s_{2}(t) \vert + \vert s_{3}(t) \vert $.\\
Therefore by using the Razumikhin-Lyapunov functional approach we obtain that the sufficient stability
condition for the synchronization manifold $x=-y$ can be written as:
\begin{equation}
\alpha > \vert m_{3} \vert +\vert m_{4} \vert + \vert K_{y} \vert
\end{equation}
As Eq.(5) is valid for small $\Delta$ stability condition (6) found above,  holds locally.
Conditions (4) are the existence conditions for the synchronization regime (3)
between bi-directionally coupled Ikeda systems (1) and (2) with multiple delays.\\
\indent At first glance, the sufficient stability condition (6) for chaos synchronization is
difficult to satisfy, as higher values of $\alpha$ could render the
dynamics trivial, i.e. non-chaotic. Fortunately there is a way out of this impasse.
The key point is that the stability conditions derived from the
Lyapunov-Razumikhin approach is a sufficient one: it assures a
high quality synchronization for a coupling strength estimated
from the stability condition, but does not forbid the possibility
of synchronization with smaller coupling strengths. Below we demonstrate
numerically that one can still achieve high quality chaos synchronization in mutually 
coupled Ikeda systems with double time delays without the condition (6) being fulfilled.\\
\indent In the case of variable time delays establishing the stability conditions for the synchronization is not as straightforward as for the constant time delays. Having in mind that for $\omega=0$ we obtain a case of constant time delays, then as an initial guess one can benefit from the existence conditions for the constant time delays case. It is our conjecture that high quality synchronization $x=-y$ could be obtained if the parameters satisfy conditions (4). As evidenced by the numerical simulations below, this conjecture is found to be well-based.\\
\indent Now consider linearly bidirectionaly coupled Ikeda systems:
\begin{equation}
\frac{dx}{dt}=-\alpha x - m_{1} \sin x_{\tau_{1}}- m_{2} \sin x_{\tau_{2}} + Ky
\end{equation}
\begin{equation}
\frac{dy}{dt}=-\alpha y - m_{3} \sin y_{\tau_{1}}- m_{4} \sin y_{\tau_{2}} + Kx
\end{equation}
For this kind of coupling the existence conditions for inverse synchronization regime (3) are:
$m_{1}=m_{3}$ and $m_{2}=m_{4};$ with sufficient stability condition:
$\alpha  > \vert m_{3} \vert +\vert m_{4} \vert + K$
\begin{center}
III. NUMERICAL SIMULATIONS
\end{center}
In numerical simulations to characterize the quality of synchronization we calculate the cross-correlation coefficient $C$ [9] 
\begin{equation}
C(\Delta t)= \frac{<(x(t) - <x>)(y(t+\Delta t) - <y>)>}{\sqrt{<(x(t) - <x>)^2><(y(t+ \Delta t) - <y>)^2>}},
\end{equation}
where $x$ and $y$ are the outputs of the interacting laser systems; the brackets$<.>$
represent the time average; $\Delta t$ is a time shift between laser outputs. This coefficient
indicates the quality of synchronization:$C=-1$ means perfect inverse synchronization.\\
\indent As mentioned above, in chaos based communication schemes synchronization between the transmitter and receiver systems are vital for message decoding. With this in mind in the remainder of the paper we focus on the inverse synchronization between bidirectionally  non-linearly and linearly coupled Ikeda systems with double variable time delays. 
Figure 1 portrays time series of the Ikeda system x (solid line) and system y (dotted line) for 
inverse chaos synchronization $x=-y$ between non-linearly coupled systems, Eqs.(1) and (2) for variable feedback time delays 
$\tau_{1}(t)=3 + 2x_{1}(t)\sin(0.15t), \tau_{2}(t)=5 + 2x_{1}(t)\sin(0.15t)$ and  variable coupling time delay $\tau_{3}(t)=7 + 2x_{1}(t)\sin(0.15t)$ with parameter values as $\alpha=3, m_{1}=m_{3}=3.1, m_{2}=m_{4}=2.5, K_{x}=K_{y}=0.03.$ $x_{1}(t)$ is the solution of Eq.(1) ($K_{y}=0$) for $\tau_{01}=3$ and $\tau_{02}=5$.
Figure 2 depicts synchronization error dynamics $\Delta =x+y$ versus time for parameters as in figure 1. C =0.99 is the cross-correlation coefficient between the transmitter $x$ and receiver $y$ system outputs. It is noted that the parameters used here in the numerical simulations satisfy the existence condition (4), but fail to satisfy the sufficient stability condition (6). Nevertheless high quality synchronization is achieved. Indeed, the value of the cross-correlation coefficient $C$ testify to the high quality chaos synchronization, which is vital for information processing in chaos-based communication systems.\\
\indent Next we present the results of numerical simulations for the linearly coupled Ikeda systems. Figure 3 presents time series of the Ikeda system x (solid line) and system y (dotted line) for 
inverse chaos synchronization $x=-y$ between linearly coupled systems, Eqs.(7) and (8) for variable feedback time delays 
$\tau_{1}(t)=1 + 0.5x_{1}(t)\sin(0.1t), \tau_{2}(t)=3 + 0.5x_{1}(t)\sin(0.1t)$ with parameter values as $\alpha=4, m_{1}=m_{3}=3.2, m_{2}=m_{4}=4.5, K=0.01.$ $x_{1}(t)$ is the solution of Eq.(8) ($K=0$) for $\tau_{01}=1$ and $\tau_{02}=3$\\
Figure 4 shows receiver output $y$ versus transmitter output $x$ for parameters as in figure 3. C =1 is the cross-correlation coefficient between the transmitter $x$ and receiver $y$ system outputs. As in the case of non-linear coupling  despite  the failure to satisfy the sufficient stability condition $\alpha  > \vert m_{3} \vert +\vert m_{4} \vert + K,$ as testified by the value of cross-correlation coefficient ($C=0.99$) we were able to achieve high quality chaos synchronization.\\ 
\begin{center}
IV. CONCLUSIONS
\end{center}
\indent To summarize we have reported on inverse chaos synchronization in bidirectionally non-linearly and linearly coupled variable multiple time delayed Ikeda systems. These results are of certain practical importance in secure chaos-based communication systems.
\begin{center}
V. ACKNOWLEDGEMENTS
\end{center}
This research was supported by a Marie Curie Action within the $6^{th}$ European Community Framework Programme Contract N:MIF2-CT-2007-039927-980065.\\
\newpage
\begin{center}
Figure Captions
\end{center}
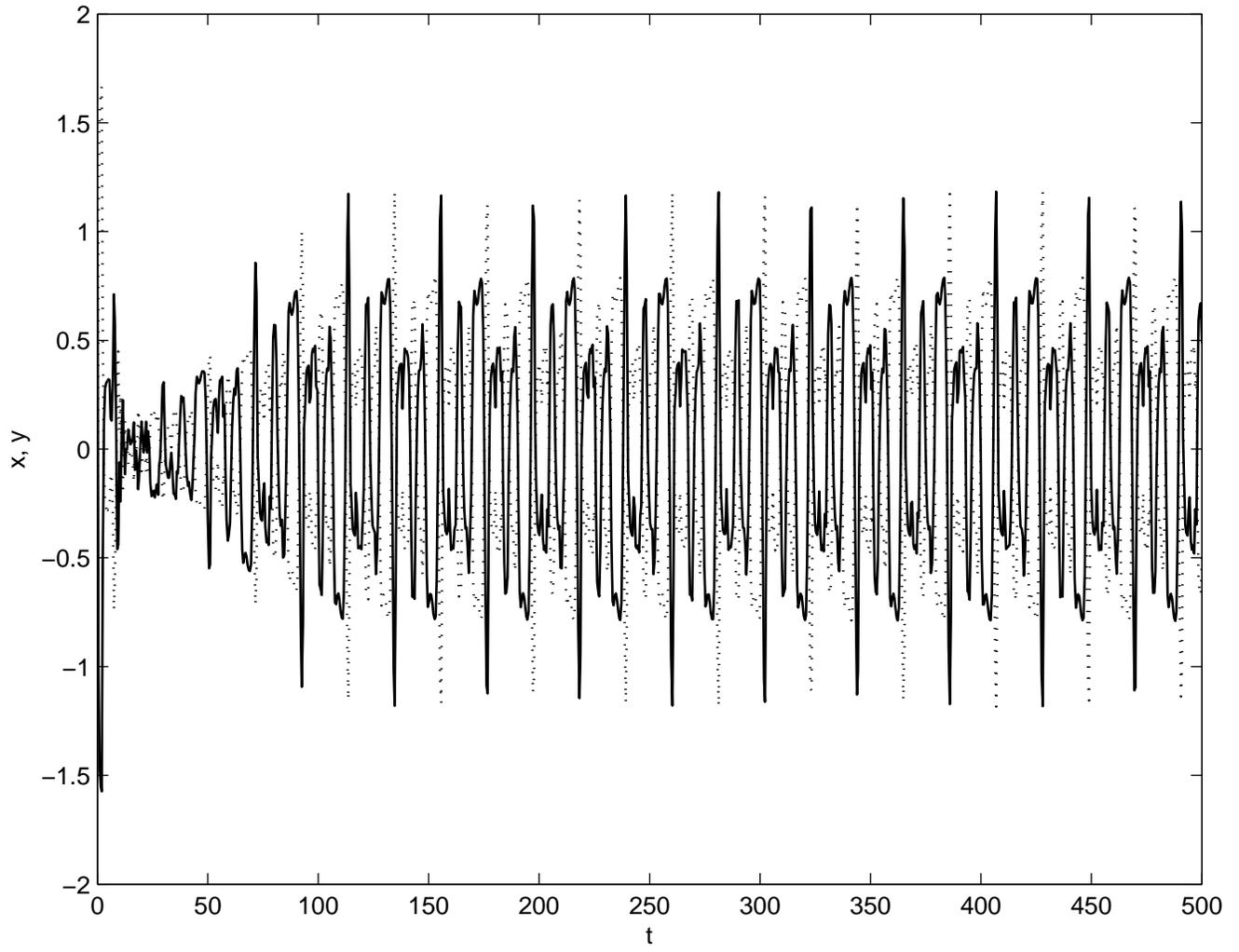
\noindent FIG.1. Numerical simulation of bidirectionally non-linearly coupled variable time delay systems, Eqs.(1-2) for $\alpha=3,m_{1}=m_{3}=3.1,m_{2}=m_{4}=2.5, K_{x}=K_{y} =0.03$ and $\tau_{1}(t)=3 + 2x_{1}(t)\sin(0.15t),\tau_{2}(t)=5 + 2x_{1}(t)\sin(0.15t),
\tau_{3}(t)=7 + 2x_{1}(t)\sin(0.15t)$. $x_{1}(t)$ is the solution of Eq.(1) ($K_{y}=0$) for $\tau_{01}=3$ and $\tau_{02}=5$.  Inverse synchronization: Time series of $x$ system (solid line) and $y$ system (dotted line). Dimensionless units.\\
~\\
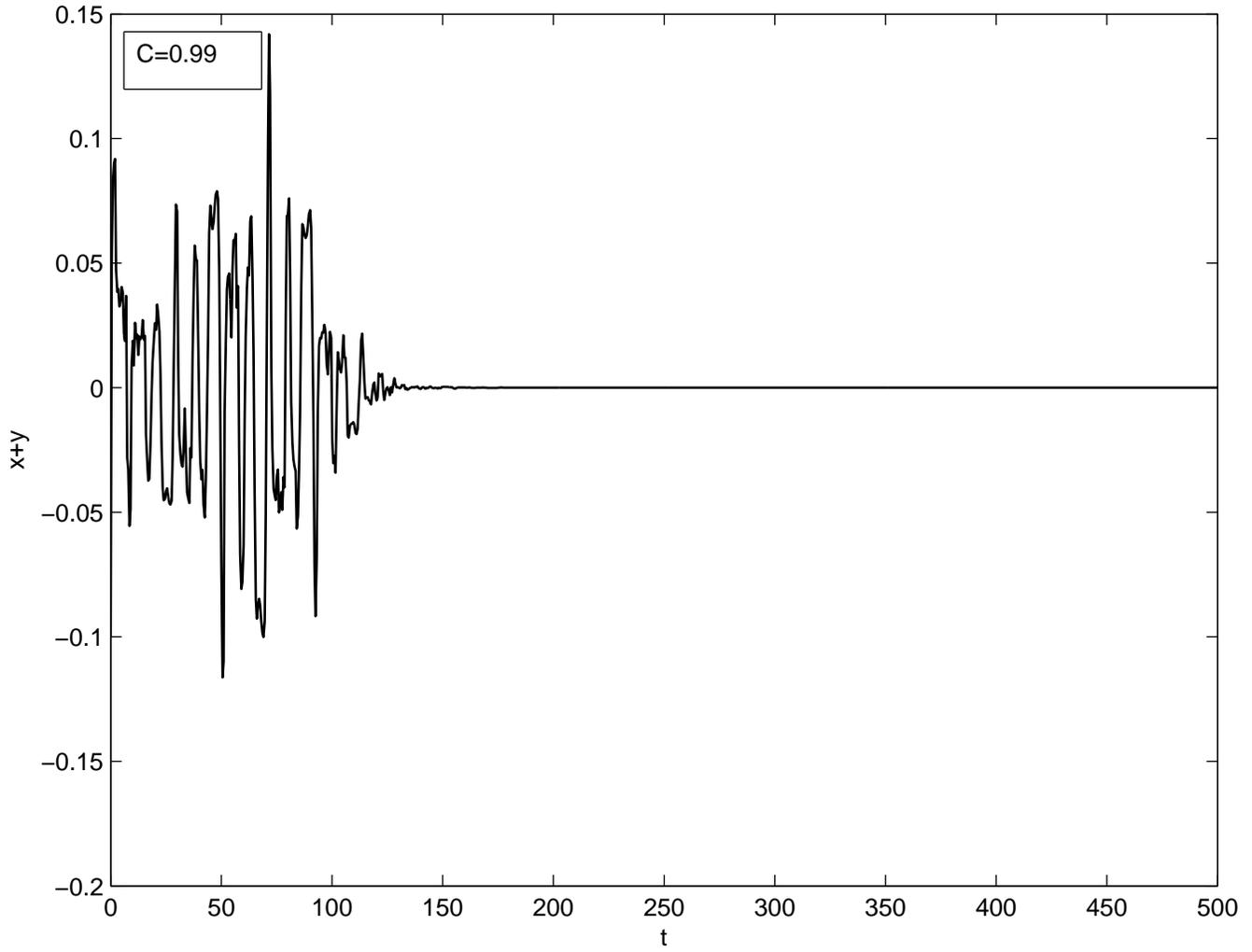
\noindent FIG.2. Numerical simulation of bidirectionally coupled variable time delay systems, Eqs.(1-2). Error dynamics, $\Delta =x+y$ versus time. The parameters are as in figure 1. C is the cross-correlation coefficient between the Ikeda systems. Dimensionless units.\\
~\\
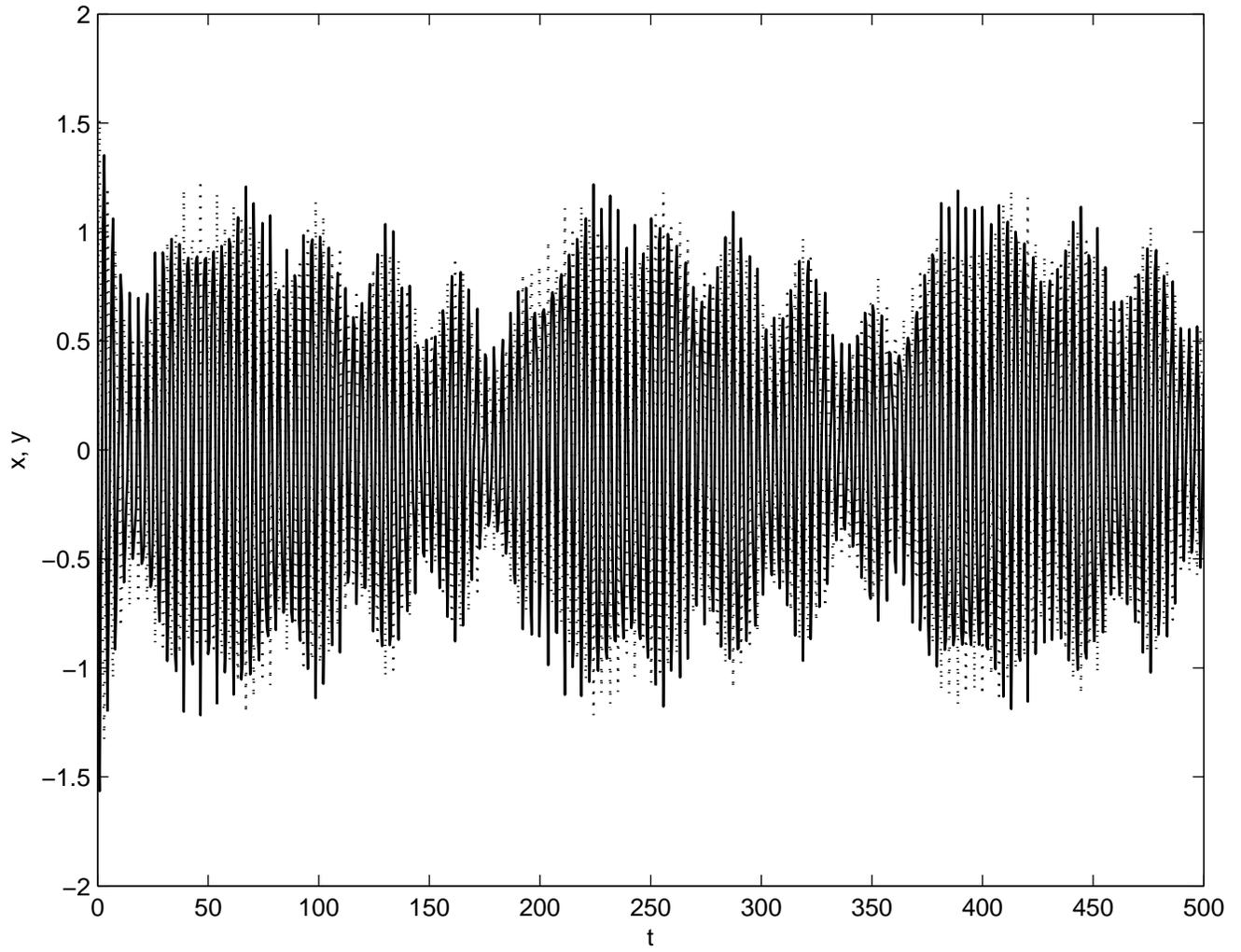
\noindent FIG.3. Numerical simulation of bidirectionally linearly coupled variable time delay systems, Eqs.(8-9) for $\alpha=4,m_{1}=m_{3}=3.2,m_{2}=m_{4}=4.5, K =0.01$ and $\tau_{1}(t)=1 + 0.5x_{1}(t)\sin(0.1t),\tau_{2}(t)=3 + 0.5x_{1}(t)\sin(0.1t)$. $x_{1}(t)$ is the solution of Eq.(8) ($K=0$) for $\tau_{01}=1$ and $\tau_{02}=3$.Inverse synchronization: Time series of $x$ system (solid line) and $y$ system (dotted line). Dimensionless units.\\
~\\
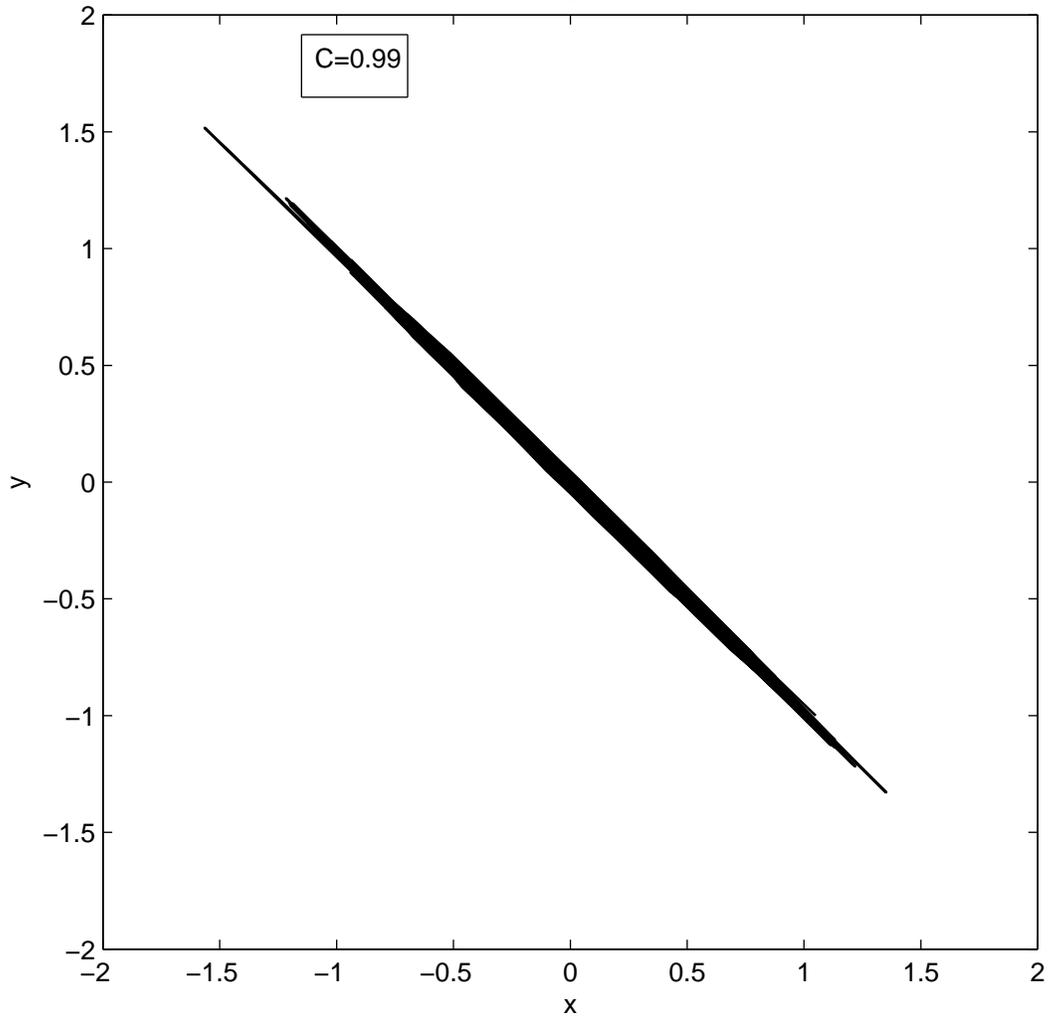
\noindent FIG.4. Numerical simulation of mutually linearly coupled  variable time delay Ikeda systems, Eqs.(7-8).Correlation plot between $x$ and $y.$ C is the correlation coefficient between $x$ and $y.$ Dimensionless units.\\

\end{document}